# Analytical solutions for five examples of shunted tunneling junctions showing promise for terahertz applications

Mark J. Hagmann

*Department of Electrical and Computer Engineering, University of Utah, 50 S. Central Campus Dr #2110, Salt Lake City, Utah 84112, USA, Correspondence: NewpathResearch@gmail.com*

**ABSTRACT**

We used analytical methods to study the interaction of electrons with shunted models consisting of a rectangular, triangular, or delta function. potential barrier in series with a pre-barrier region at zero potential. In each model the shunted boundary conditions cause the matrix equation to have only zeros in the right-hand column vector. Thus, the determinant for each matrix must be zero for a non-trivial solution. The determinant for each model contains only the parameters (e.g. the length of the shunt, the length and height of the barrier, and the electron energy). Thus, the complete set of solutions for each model is obtained by using algebra to determine all of the points in the parameter space, and then to calculate the coefficients for each model. Any path from one point to another in the parameter space corresponds to a possible history for the operation of the model. In prototypes the shunts could be filaments of certain metals that provide quasi-coherent electron transport over mean-free paths of tens of nm. Quasi-static simulations of the time-independent Schrödinger equation suggest that a device with a size of 100 nm could operate at frequencies up to 1,000 THz.

. **I. INTRODUCTION**

We present analytical solutions of the Schrödinger equation with models of closed quantum nanocircuits. Each model consists of a tunneling junction shunted by a filament of a wire such as beryllium in which electrons have a mean-free path as great as 68 nm to approximate coherent transport [1],[2]. Thus, we consider transport in the wire as though it were vacuum. Cases for a nanowire with zero length are considered separately. The two points that are connected to close each circuit are marked by ground symbols and cases for which the wire filament has zero length are considered separately. The system of equations that is solved for each model has only homogeneous equations so the solution is determined by requiring that the determinant of the corresponding matrix is zero. A step-size was chosen for the curves in each figure but they would become a continuum as the step-size approaches zero.

In quantum mechanics generally two terms with imaginary exponentials are required in the pre-barrier region, which is at zero potential, to represent a forward and a backward wave [3]. However, now with the shunted models there are only standing waves in that region so we use sine and cosine terms and no complex arithmetic is required in the analysis.

**II. MODEL 1: CLOSED-CIRCUIT WITH A RECTANGULAR BARRIER**

Figure 1 shows the potential for a one-dimensional model having a rectangular barrier with a pre-barrier region that is at zero potential. Note that the parameter a is negative. The two lower ends of this model are marked with ground symbols so show that they are connected at a single point to complete the circuit. Thus, this model represents a ring with an opening as Region II.

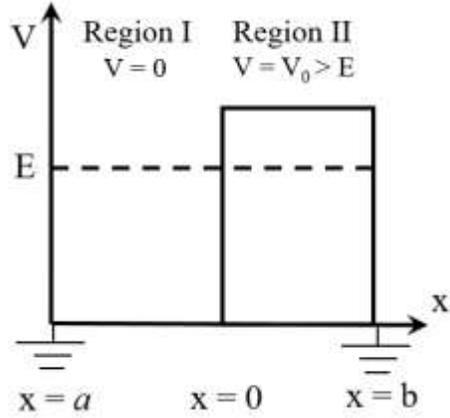

Fig. 1. Potential energy for a closed circuit with a rectangular barrier.

We write the time-independent Schrödinger equation as Eq. 1, where E is the energy of the particle and V(x) is the potential energy.

$$\frac{d^2\psi}{dx^2} + \frac{2m}{\hbar^2}[E - V(x)]\psi = 0 \tag{1}$$

Solving the Schrödinger equation, we obtain Equations 2 and 3 for the wavefunctions in regions I and II where the parameters k and β are defined in Eqns. 4 and 5. Taking the derivatives in respect to x gives Eqns. 6 and 7.

$$\psi_I = A\cos(kx) + B\sin(kx) \tag{2}$$

$$\psi_{II} = Ce^{\beta x} + De^{-\beta x} \tag{3}$$

$$k \equiv \frac{\sqrt{2mE}}{\hbar} \tag{4}$$

$$\beta \equiv \frac{\sqrt{2m(V_0 - E)}}{\hbar} \tag{5}$$

$$\frac{d\psi_I}{ds} = -kA\sin(kx) + kB\cos(kx) \tag{6}$$

$$\frac{d\psi_{II}}{dx} = \beta Ce^{\beta x} - \beta De^{-\beta x} \tag{7}$$

Applying the boundary conditions so that the wavefunction and its derivative are continuous for x equal to zero gives Eqns. 8 and 9.

$$A - C - D = 0 \tag{8}$$

$$kB - \beta C + \beta D = 0 \tag{9}$$

Requiring continuity of the wave function and its derivative at the common boundary where x equals a in region I to that where x equals b in region II gives Eqns. 10 and 11.

$$A\cos(ka) + B\sin(ka) - Ce^{\beta b} - De^{-\beta b} = 0 \tag{10}$$

$$kA\sin(ka) - kB\cos(ka) + \beta Ce^{\beta b} - \beta De^{-\beta b} = 0 \tag{11}$$

The system of Eqs, 8, 9, 10, and 11 in the four coefficients A, B, C, and D may be written in matrix form as shown in Eq. 12.

$$\begin{bmatrix} +1 & 0 & -1 & -1 \\ 0 & +k & -\beta & +\beta \\ +\cos(ka) & +\sin(ka) & -e^{\beta b} & -e^{-\beta b} \\ +k\sin(ka) & -k\cos(ka) & +\beta e^{\beta b} & -\beta e^{-\beta b} \end{bmatrix} \begin{bmatrix} A \\ B \\ C \\ D \end{bmatrix} = \begin{bmatrix} 0 \\ 0 \\ 0 \\ 0 \end{bmatrix} \quad (12)$$

However, this is a homogeneous system of equations so to have a non-trivial solution for the four coefficients the determinant must be zero, as shown in Eq. 13.

$$\begin{vmatrix} +1 & 0 & -1 & -1 \\ 0 & +k & -\beta & +\beta \\ +\cos(ka) & +\sin(ka) & -e^{\beta b} & -e^{-\beta b} \\ +k\sin(ka) & -k\cos(ka) & +\beta e^{\beta b} & -\beta e^{-\beta b} \end{vmatrix} = 0 \quad (13)$$

In general, expanding a determinant with four rows and four columns will give 24 terms. However, for the determinant in Eq. 13, ten terms are zero to reduce this to 14 terms. Also setting the matrix elements that are plus or minus 1 in these 14 terms gives the simpler expression in Eq. 14 for the determinant.

$$Det = -M_{23}M_{32}M_{41} + M_{24}M_{32}M_{41} + M_{22}M_{33}M_{41} - M_{22}M_{34}M_{41} + M_{23}M_{31}M_{42}$$
$$- M_{24}M_{31}M_{42} - M_{24}M_{33}M_{42} + M_{23}M_{34}M_{42} - M_{22}M_{31}M_{43} + M_{24}M_{32}M_{43}$$
$$- M_{22}M_{34}M_{43} + M_{22}M_{31}M_{44} - M_{23}M_{32}M_{44} + M_{22}M_{33}M_{44} \quad (14)$$

Entering the expressions for each of the remaining terms in Eq. 14, introducing hyperbolic functions, and simplifying gives Eq. 15. For the system of equations to have a non-zero solution it is necessary for the determinant to be zero as shown in Eq. 16.

$$Det = 2(\beta^2 - k^2)\sinh(\beta b)\sin(ka) + 4k\beta[1 - \cosh(\beta b)\cos(ka)] \quad (15)$$

To obtain the solution, first we specify values for the energy E and the potential $V_0$. Then we use Eq. 4 and Eq. 5 to determine k and β. Next, we iterate to determine the two remaining parameters, a and b, such that the determinant is zero, as shown in Eq. 16, to obtain unique solutions.

$$(\beta^2 - k^2)\sinh(\beta b)\sin(ka) + 2k\beta[1 - \cosh(\beta b)\cos(ka)] = 0 \quad (16)$$

When any three of the four parameters a, b, E, and $V_0$ are specified the determinant may be brought to zero by changing the fourth parameter. Figure 2 shows the values of a and E when $V_0$ is 1 Volt for different values of b.

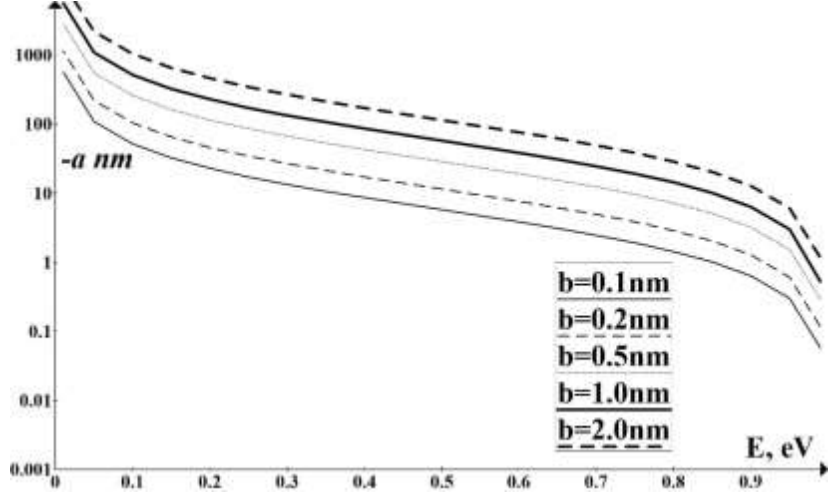

Fig. 2. Solutions for a rectangular barrier with the parameters a and E for different lengths of the barrier b where $V_0$ is 1 volt.

For the special case of a shorted rectangular barrier the constant "a" must be zero. Thus, the first term in Eq. 16 is zero to give Eq. 17. Thus, either E equals $V_0$ or else b is zero, where in either case there would be no tunneling. Thus, there would only be trivial solutions in which there is no barrier or pre-barrier for the model.

$$2k\beta\left[1-\cosh(\beta b)\right]=0 \quad (17)$$

### III. COEFICIENTS FOR THE RECTANGULAR BARRIER WAVE FUNCTION

First, we normalize the system of four equations by setting A to be unity in Eq. 18 and dividing Eqns. 9 and 11 by k to form Eqns. 20, 21, 22, and 23.

$$A=1 \quad (18)$$
$$C+D=1 \quad (19)$$
$$B-\frac{\beta}{k}C+\frac{\beta}{k}D=0 \quad (20)$$
$$\cos(ka)+B\sin(ka)-Ce^{\beta b}-De^{-\beta b}=0 \quad (21)$$
$$\sin(ka)-B\cos(ka)+\frac{\beta}{k}Ce^{\beta b}-\frac{\beta}{k}De^{-\beta b}=0 \quad (22)$$

Combining Eqns. 19 and 20 we obtain the following solutions for C and D:

$$C=\frac{1}{2}\left(1+\frac{k}{\beta}\right) \quad (23)$$
$$D=\frac{1}{2}\left(1-\frac{k}{\beta}\right) \quad (24)$$

Finally, substituting Eqns. 23 and 24 into Eq. 22 gives Eq. 25 for the coefficient B.

$$B = \frac{\beta}{2k\cos(ka)}\left(1+\frac{k}{\beta}\right)e^{\beta b} - \frac{\beta}{2k\cos(ka)}\left(1-\frac{k}{\beta}\right)e^{-\beta b} + \tan(ka) \quad (25)$$

## IV. MODEL 2: CLOSED-CIRCUIT WITH A TRIANGULAR BARRIER

Landau and Lifshitz [4] presented an exact solution of the Schrödinger equation for the one-dimensional motion of a particle in a homogeneous external field in terms of Airy functions that was first published in 1965. Now we present a solution for a triangular barrier having finite linear extent so it is necessary to include both Airy functions with their derivatives in the solution. It is also necessary to close the circuit model as shown in Fig. 3. As in Fig. 1, the two ground symbols represent a common connection with zero length to complete the closed circuit. Note that the potential $V_0$ has units of energy and the coordinate labeled "a" has a negative value. The left edge of the barrier represents an ideal voltage source.

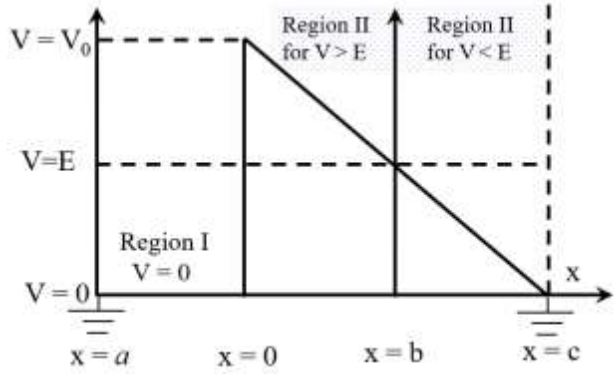

Fig. 3. Potential energy in regions I and II with a triangular barrier

To keep the derivations for the first and second models separate we use Eq. 27 for the wavefunction in Region I with the definition of k in Eq. 4. The coefficients A and B are constants to be determined by the boundary conditions.

$$\psi_I = A\cos(kx) + B\sin(kx) \quad (27)$$

In Region II the time-independent Schrödinger equation is written as Eq. 28 where E and V have units of electron volts and e is the magnitude of the charge on an electron.

$$\frac{d^2\psi_{II}}{dx^2} + \frac{2me}{\hbar^2}[E-V(x)]\psi_{II} = 0 \quad (28)$$

The potential energy within the barrier is given by Eq. 29 as a linear interpolant. Thus, the x coordinate b at which the potential energy is equal to the particle energy is given by Eq. 30.

$$V(x) = \left(1-\frac{x}{c}\right)V_0 \quad (29)$$

$$b = \left(1-\frac{E}{V_0}\right)c \quad (30)$$

Substituting the expression for the potential energy from Eq. 29 into Eq. 28 gives Eq. 31.

$$\frac{d^2\psi_{II}}{dx^2} + \frac{2me}{\hbar^2}\left[(E-V_0)+V_0\frac{x}{c}\right]\psi_{II} = 0 \quad (31)$$

Next, we make a change in the parameters to derive a solution in terms of Airy functions. This method is presented in the Wolfram Alpha differential equation solver. Here we make a change of the parameters of Eq. 31 to obtain Eq. 32 where the new argument $\xi$ is defined in Eq. 33. The coefficients C and D are determined using the boundary conditions.

$$\psi_{II}(x) = CAi(\xi) + DBi(\xi) \tag{32}$$

$$\xi \equiv \left(\frac{2mV_0e}{\hbar^2 c}\right)^{\frac{1}{3}} \left(\frac{E}{V_0} - 1\right) c + \left(\frac{2mV_0e}{\hbar^2 c}\right)^{\frac{1}{3}} x \tag{33}$$

We separate the argument $\xi$ into two parts as shown in Eq. 34 where $\gamma$ and K are defined in Eqns. 35 and 36. Thus, the wavefunction and its derivative in Region II are given by Eqns. 37 and 38. Note that $\gamma$ is greater than zero, and K is negative for E less than $V_0$ in quantum tunneling.

$$\xi = K + \gamma x \tag{34}$$

$$\gamma \equiv \left(\frac{2mV_0e}{\hbar^2 c}\right)^{\frac{1}{3}} \tag{35}$$

$$K \equiv -\left(1 - \frac{E}{V_0}\right)\gamma c \tag{36}$$

$$\psi_{II}(x) = CAi(K + \gamma x) + DBi(K + \gamma x) \tag{37}$$

$$\psi_{II}'(x) = \gamma CAi'(K + \gamma x) + \gamma DBi'(K + \gamma x) \tag{38}$$

From Eq 27, in Region I the derivative of the wavefunction is given by Eq. 39.

$$\psi_I'(x) = -kA\sin(kx) + kB\cos(kx) \tag{39}$$

Next, we apply the boundary conditions for the closed-circuit. First, the wavefunctions and their derivatives just inside the two ends of Region I are given by Eqns. 40 to 43.

$$\psi_I(0^-) = A \tag{40}$$

$$\psi_I'(0^-) = kB \tag{41}$$

$$\psi_I(a^+) = \cos(ka)A + \sin(ka)B \tag{42}$$

$$\psi_I'(a^+) = -k\sin(ka)A + k\cos(ka)B \tag{43}$$

The wavefunctions and their derivatives just inside the two ends of Region II are given by Eqns. 44 to 47.

$$\psi_{II}(0^+) = Ai(K)C + Bi(K)D \tag{44}$$

$$\psi_{II}'(0^+) = \gamma Ai'(K)C + \gamma Bi'(K)D \tag{45}$$

$$\psi_{II}(c^-) = Ai(K + \gamma c)C + Bi(K + \gamma c)D \tag{46}$$

$$\psi_{II}'(c^-) = \gamma Ai'(K + \gamma c)C + \gamma Bi'(K + \gamma c)D \tag{47}$$

Next pairs from the system of Eqns. 40 to 47 are used to form the following system of 4 simultaneous equations in the 4 unknown coefficients as required for continuity of the wavefunctions and their derivatives.
From Eqns. 40 and 44, for $\psi_I(0^-)$ equal to $\psi_{II}(0^+)$:

$$A = Ai(K)C + Bi(K)D \tag{48}$$

From Eqns. 41 and 45, for $\psi_I'$ (0⁻) equal to $\psi_{II}'$ (0⁺):

$$kB = \gamma Ai'(K)C + \gamma Bi'(K)D \tag{49}$$

From Eqns. 42 and 46, for $\psi_I$ (a⁺) equal to $\psi_{II}$ (c⁻):

$$\cos(ka)A + \sin(ka)B = Ai(K+\gamma c)C + Bi(K+\gamma c)D \tag{50}$$

From Eqns. 43 and 47, for $\psi_I'$ (a⁺) equal to $\psi_{II}'$(c⁻):

$$-k\sin(ka)A + k\cos(ka)B = \gamma Ai'(K+\gamma c)C + \gamma Bi'(K+\gamma c)D \tag{51}$$

Next, we rearrange Eqns. 48 to 51 to have zero on the right-hand side and divide those with $\gamma$ by k and use the parameter R as defined in Eqn. 52 to obtain Eqns. 53 to 56.

$$R \equiv \frac{\gamma}{k} \tag{52}$$

$$A - Ai(K)C - Bi(K)D = 0 \tag{53}$$

$$B - RAi'(K)C - RBi'(K)D = 0 \tag{54}$$

$$\cos(ka)A + \sin(ka)B - Ai(K+\gamma c)C - Bi(K+\gamma c)D = 0 \tag{55}$$

$$\sin(ka)A - \cos(ka)B + RAi'(K+\gamma c)C + RBi'(K+\gamma c)D = 0 \tag{56}$$

Next, we write this system of four simultaneous homogeneous equations as a matrix equation in Eq. 57.

$$\begin{bmatrix} 1 & 0 & -Ai(K) & -Bi(K) \\ 0 & 1 & -RAi'(K) & -RBi'(K) \\ \cos(ka) & \sin(ka) & -Ai(K+\gamma c) & -Bi(K+\gamma c) \\ \sin(ka) & -\cos(ka) & +RAi'(K+\gamma c) & +RBi'(K+\gamma c) \end{bmatrix} \begin{bmatrix} A \\ B \\ C \\ D \end{bmatrix} = \begin{bmatrix} 0 \\ 0 \\ 0 \\ 0 \end{bmatrix} \tag{57}$$

To simplify the notation in this matrix equation, we define the parameters Θ and X, where Θ is less than zero with units of radians. Then Eq. 36 is combined with Eq. 59 to obtain Eq. 60 which shows that X is greater than zero. Finally, Eqns. 57, 58, and 59 are combined to obtain Eq. 61, where a is less than zero so Θ is also less than zero.

$$\Theta \equiv ka \tag{58}$$

$$X \equiv K + \gamma c \tag{59}$$

$$X = \gamma c \frac{E}{V_0} \tag{60}$$

$$\begin{bmatrix} 1 & 0 & -Ai(K) & -Bi(K) \\ 0 & 1 & -RAi'(K) & -RBi'(K) \\ \cos(\Theta) & \sin(\Theta) & -Ai(X) & -Bi(X) \\ \sin(\Theta) & -\cos(\Theta) & +RAi'(X) & +RBi'(X) \end{bmatrix} \begin{bmatrix} A \\ B \\ C \\ D \end{bmatrix} = \begin{bmatrix} 0 \\ 0 \\ 0 \\ 0 \end{bmatrix} \tag{61}$$

Again, note that X is greater than zero, and K is less than zero when E is less than $V_0$ to permit quantum tunneling. The matrix equation defined by Eq. 61 is homogeneous so the determinant of the matrix must be zero, as shown explicitly in Eq. 62 to have a non-trivial solution for the coefficients. Thus, it is shown that the solutions may be determined by varying the parameters R, K, X and Θ to bring the determinant to zero.

$$Det = \begin{vmatrix} 1 & 0 & -Ai(K) & -Bi(K) \\ 0 & 1 & -RAi'(K) & -RBi'(K) \\ \cos(\Theta) & \sin(\Theta) & -Ai(X) & -Bi(X) \\ \sin(\Theta) & -\cos(\Theta) & +RAi'(X) & +RBi'(X) \end{vmatrix} = 0 \quad (62)$$

There are 24 terms in the determinant, but 10 of them are zero leaving the following 14 in Eq. 63.

$$Det = M_{11}M_{22}M_{33}M_{44} + M_{11}M_{23}M_{34}M_{42} + M_{11}M_{24}M_{32}M_{43} - M_{11}M_{24}M_{33}M_{42}$$
$$- M_{11}M_{23}M_{32}M_{44} - M_{11}M_{22}M_{34}M_{43} + M_{13}M_{24}M_{31}M_{42} + M_{14}M_{22}M_{31}M_{43}$$
$$- M_{14}M_{23}M_{31}M_{42} - M_{13}M_{22}M_{31}M_{44} - M_{13}M_{24}M_{32}M_{41} - M_{14}M_{22}M_{33}M_{41}$$
$$+ M_{14}M_{23}M_{32}M_{41} + M_{13}M_{22}M_{34}M_{41} \quad (63)$$

The terms with indices 11 and 22 in the determinant are both equal to 1 which simplifies the 14 terms as shown in Eq. 64

$$Det = M_{33}M_{44} + M_{23}M_{34}M_{42} + M_{24}M_{32}M_{43} - M_{24}M_{33}M_{42} - M_{23}M_{32}M_{44}$$
$$- M_{34}M_{43} + M_{13}M_{24}M_{31}M_{42} + M_{14}M_{31}M_{43} - M_{14}M_{23}M_{31}M_{42}$$
$$- M_{13}M_{31}M_{44} - M_{13}M_{24}M_{32}M_{41} - M_{14}M_{33}M_{41} + M_{14}M_{23}M_{32}M_{41}$$
$$+ M_{13}M_{34}M_{41} \quad (64)$$

Inserting the expressions for each term from Eq. 62 into Eq. 64 gives the following expression for the determinant as Eq. 65.

$$Det = -RAi(X)Bi'(X) - RAi'(K)Bi(X)\cos(\Theta) - R^2Ai'(X)Bi'(K)\sin(\Theta)$$
$$+ RAi(X)Bi'(K)\cos(\Theta) + R^2Ai'(K)Bi'(X)\sin(\Theta) + RAi'(X)Bi(X)$$
$$- RAi(K)Bi'(K)\cos^2(\Theta) - RAi'(X)Bi(K)\cos(\Theta) + RAi'(K)Bi(K)\cos^2(\Theta)$$
$$+ RAi(K)Bi'(X)\cos(\Theta) - RAi(K)Bi'(K)\sin^2(\Theta) - Ai(X)Bi(K)\sin(\Theta)$$
$$+ RAi'(K)Bi(K)\sin^2(\Theta) + Ai(K)Bi(X)\sin(\Theta) \quad (65)$$

Combining the squares of the sine and cosine terms reduces the number of terms from 14 to 12 in Eq. 66.

$$Det = +R^2Ai'(K)Bi'(X)\sin(\Theta) - R^2Ai'(X)Bi'(K)\sin(\Theta)$$
$$+ RAi'(X)Bi(X) - RAi(X)Bi'(X)$$
$$+ RAi'(K)Bi(K) - RAi(K)Bi'(K)$$
$$+ RAi(X)Bi'(K)\cos(\Theta) - RAi'(X)Bi(K)\cos(\Theta)$$
$$+ RAi(K)Bi'(X)\cos(\Theta) - RAi'(K)Bi(X)\cos(\Theta)$$
$$+ Ai(K)Bi(X)\sin(\Theta) - Ai(X)Bi(K)\sin(\Theta) \quad (66)$$

Finally, factoring Eq. 66 gives the more functional form in Eq. 67.

$$Det = +R^2\left[Ai'(K)Bi'(X) - Ai'(X)Bi'(K)\right]\sin(\Theta)$$
$$+ R\left[Ai'(X)Bi(X) - Ai(X)Bi'(X)\right]$$
$$+ R\left[Ai'(K)Bi(K) - Ai(K)Bi'(K)\right]$$
$$+ R\left[Ai(X)Bi'(K) - Ai'(X)Bi(K)\right]\cos(\Theta)$$
$$+ R\left[Ai(K)Bi'(X) - Ai'(K)Bi(X)\right]\cos(\Theta)$$
$$+ \left[Ai(K)Bi(X) - Ai(X)Bi(K)\right]\sin(\Theta) \qquad (67)$$

We find the null-points for the determinant as follows:
1. Input values for the energy E, the potential $V_0$, and the barrier length c.
2. Equation 68 may be used to determine b, the length for tunneling within the triangular barrier but this quantity is not required in the calculations.

$$b = \left(1 - \frac{E}{V_0}\right)c \qquad (68)$$

3. Use Eqns. 4, 36, 35 and 52 to determine the corresponding values of k, K, γ, and R.
4. Form the determinant and vary Θ to bring the determinant to zero.
5. We may only use negative values of Θ and positive values of k in Eqns. 58 and 4 because $a$, the coordinate at the left end of the model, is less than zero.

This algorithm was used to generate the data that are plotted in Fig. 4. Note that the vertical distance between the two curves is extremely small. A least-squares linear regression of these data shows that the angle Θ in degrees has a mean value of -359.77003 with a standard deviation of 0.178557 in degrees. These calculations were made using 99 values of the particle energy from 0.01 to 0.99 electron volts for higher precision in the least-squares linear regression.

This algorithm may be extended to make plots of E or $V_0$ as functions of the parameters a and c as follows:
1. Input the values of E, $V_0$, and c.
2. Use E in Eq. 4 to determine the corresponding values of k.
3. Use k and Θ to determine the pre-barrier length a.
4. Then plot E and/or $V_0$ as a function of two parameters a and c.

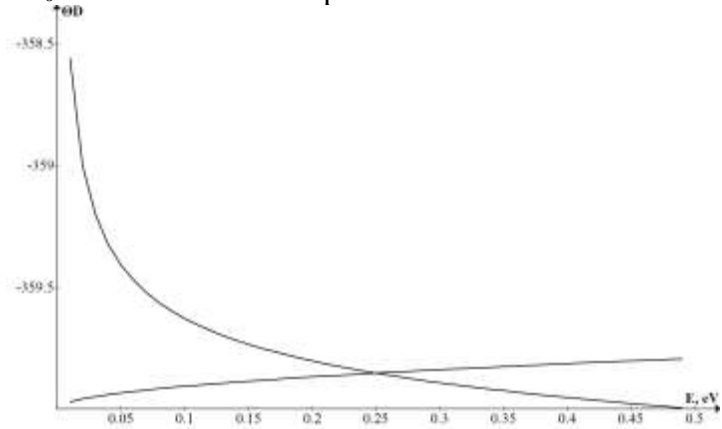

Fig 4. Two roots for Θ as a function of the particle energy E with a triangular barrier having b = 0.1 nm and c = 1 nm with $V_0$ = 1.0 V.

Figure 5 shows the parameters a and c for a potential $V_0$ of 1 V at specified values of the energy E. There are two sets of data, and in each set the energy goes from 0.99 eV at the top to 0.01 eV at the base in steps of 0.01 eV. Note that the magnitude of the step size in a increases monotonically from the top to the bottom in each of the two sets. The overlap in these two sets causes multiple values of E for values of a and c near where these two sets cross. These two sets merge as the step size is made much smaller than the value of 0.01 eV.

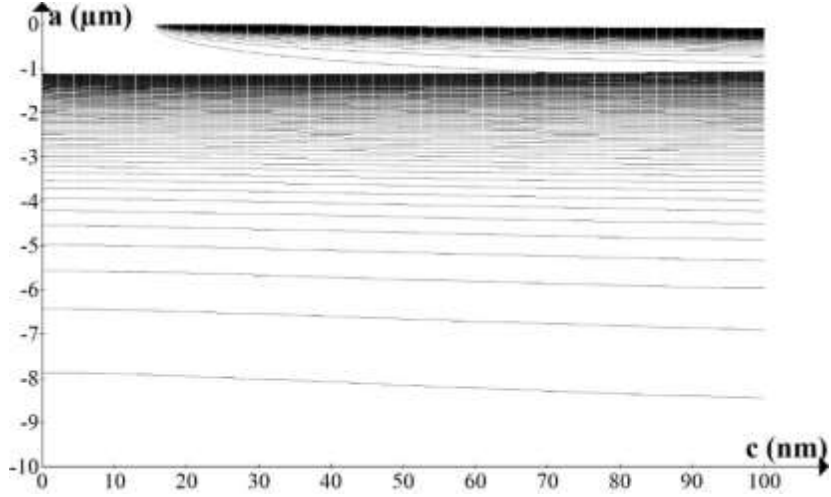

Fig. 5. Two overlapping sets for the parameters a and c with a potential of 1 V at specific values of the energy in steps of 0.01 eV.

## V. COEFFICIENTS FOR THE TRIANGULAR BARRIER WAVEFUNCTION.

We have shown that the determinant which is defined in Eq. 62 must be zero. Now we copy the four component equations from Eq. 61 as Eqs. 69, 70, 71 and 72, which are normalized by defining the coefficient A to be unity, and then dividing Eqns. 70 and 72 by γ.

$$1 - Ai(K - \gamma c)C - Bi(K - \gamma c)D = 0 \tag{69}$$

$$\frac{k}{\gamma}B + Ai'(K - \gamma c)C + Bi'(K - \gamma c)D = 0 \tag{70}$$

$$\cos(ka) + \sin(ka)B - Ai(K - \gamma a)C - Bi(K - \gamma a)D = 0 \tag{71}$$

$$-\frac{k}{\gamma}\sin(ka) + \frac{k}{\gamma}\cos(ka)B + Ai'(K - \gamma a)C + Bi'(K - \gamma a)D = 0 \tag{72}$$

Any three of these four equations may be combined to solve for the normalized coefficients. For example, we may combine Eqns. 70 and 71 to obtain Eq. 73, and use Eq. 69 to determine C and D, after which Eq. 73 is used to obtain the normalized coefficient B.

$$\left[\frac{\gamma}{k}Ai'(K-\gamma c)\sin(ka) + Ai(K-\gamma a)\right]C + \left[\frac{\gamma}{k}Bi'(K-\gamma c)\sin(ka) + Bi(K-\gamma a)\right]D = -\cos(ka) \tag{73}$$

# VI. MODEL 3: SPECIAL CASE FOR A SHORTED TRIANGULAR BARRIER

Figure 6 shows the special case in which the pre-barrier region is deleted to form a closed circuit having only the shorted triangular barrier. This model may be interpreted as being a tunneling junction shunted by a source of electrical potential which has zero length.

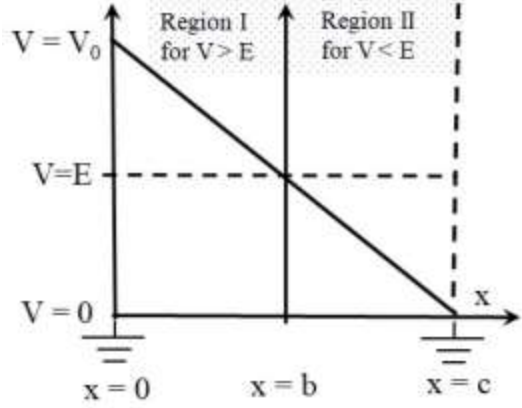

Fig. 6. Potential energy in shorted triangular barrier

Now, because $\Theta$ is zero with no pre-barrier, the twelve terms for the determinant in Eq. 66 are reduced to the 8 terms in Eq. 74.

$$Det = \begin{bmatrix} Ai'(X)Bi(X) - Ai(X)Bi'(X) + Ai'(K)Bi(K) - Ai(K)Bi'(K) \\ + Ai(X)Bi'(K) - Ai'(X)Bi(K) + Ai(K)Bi'(X) - Ai'(K)Bi(X) \end{bmatrix} R \quad (74)$$

Figure 6, shows the roots for which, the determinant is zero as a function of γc and $E/V_0$. It was prepared by determining the roots for specific values for X and K, and then use Eqns. 76 and 77 to determine the corresponding values of γc and $E/V_0$.

$$Det = \begin{bmatrix} Ai'\left[\frac{E}{V_0}\gamma c\right]Bi\left[\frac{E}{V_0}\gamma c\right] - Ai\left[\frac{E}{V_0}\gamma c\right]Bi'\left[\frac{E}{V_0}\gamma c\right] \\ + Ai'\left[-\left(1-\frac{E}{V_0}\right)\gamma c\right]Bi\left[-\left(1-\frac{E}{V_0}\right)\gamma c\right] - Ai\left[-\left(1-\frac{E}{V_0}\right)\gamma c\right]Bi'\left[-\left(1-\frac{E}{V_0}\right)\gamma c\right] \\ + Ai\left[\frac{E}{V_0}\gamma c\right]Bi'\left[-\left(1-\frac{E}{V_0}\right)\gamma c\right] - Ai'\left[\frac{E}{V_0}\gamma c\right]Bi\left[-\left(1-\frac{E}{V_0}\right)\gamma c\right] \\ + Ai\left[-\left(1-\frac{E}{V_0}\right)\gamma c\right]Bi'\left[\frac{E}{V_0}\gamma c\right] - Ai'\left[-\left(1-\frac{E}{V_0}\right)\gamma c\right]Bi\left[\frac{E}{V_0}\gamma c\right] \end{bmatrix} R \quad (75)$$

$$\frac{V_0}{E} = 1 - \frac{K}{X} \quad (76)$$

$$\gamma c = X - K \qquad (77)$$

Figure 7 shows the roots where the determinant in Eq. 75 is zero as a function of two parameters: the energy E divided by the potential $V_0$ and the product γc.

At each point in Fig. 7 the potential is determined by using the following procedure:
1. Choose any point on the curve in Fig. 7, and note the specific values for the pair of the dimensionless parameters E/$V_0$ and γc.
2. Specify values for the mass of the electron or other chosen particle.
3. Specify the value for c, which is the total length of the triangular barrier.
4. Divide the value for γc at the chosen point by c to determine γ.
5. Equation 78, which was derived from Eq. 35, is used to solve for the value of the potential $V_0$ in volts to complete the solution at the chosen point.

$$V_0 = \frac{\hbar^2 \gamma^3 c}{2me} \qquad (78)$$

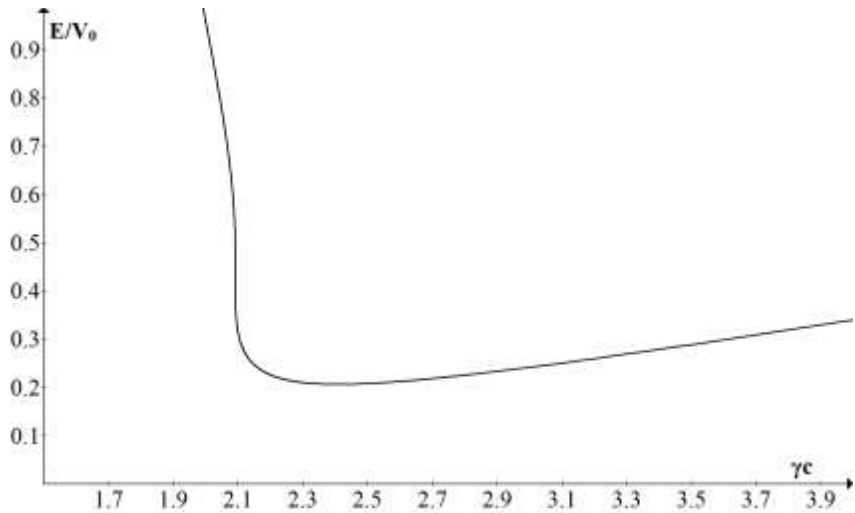

Fig. 7 Roots for the determinant in Eq. 75 to be zero as a continuous function of the two parameters E/$V_0$ and γc for the shorted triangular barrier.

Figure 8 shows the results from a second study using Eq. 75 for the determinant with a shorted triangular barrier. Note that the energy required for tunneling increases monotonically as a function of the product of the length of the barrier and the potential.

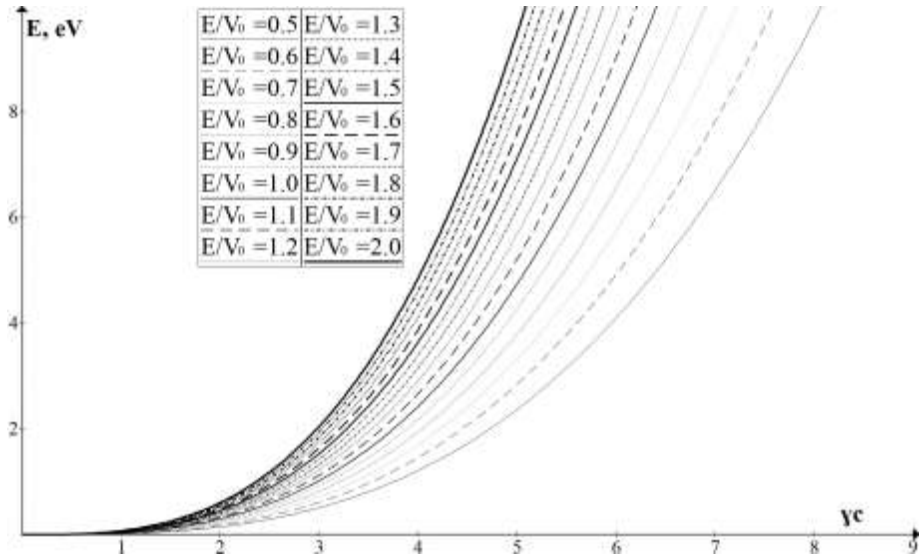

Fig. 8. Energy E as a function of the product γc for a shorted triangular barrier with a length c of 1 nanometer for 16 different ratios of the energy to the potential.

Figure 9 shows that there is a transition between Region I and Region II at the point where x equals b at which the energy is equal to the potential. This transition occurs at x equal to zero for the special case where the energy is equal to $V_0$, and at x equal to c when the energy is zero. For intermediate values of the energy this transition occurs as shown in Eq. 79. Note the linear decrease in the parameter b as the energy is increased, which would be expected.

$$b = \left(1 - \frac{E}{V_0}\right)c \qquad (79)$$

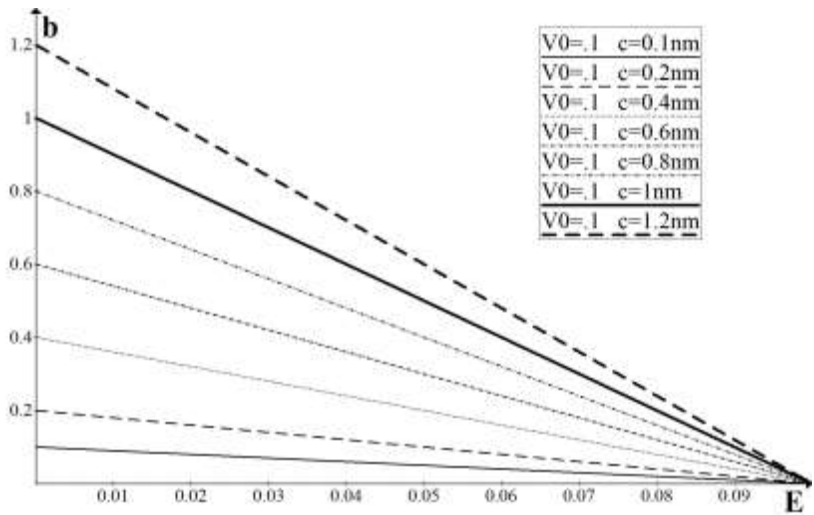

Fig. 9. The value of the parameter b at the transition between quantum tunneling and classical transport with the shorted triangular barrier as a function of the energy and the potential.

## VII. MODEL 4: CLOSED-CIRCUIT WITH A DELTA-FUNCTION BARRIER

Figure 10 shows a closed nano-circuit with a delta-function barrier at the origin and a potential of zero elsewhere. As in the previous sections of this paper, the two ground symbols denote a connection so that "x" represents the coordinate at all the points in the closed circuit.

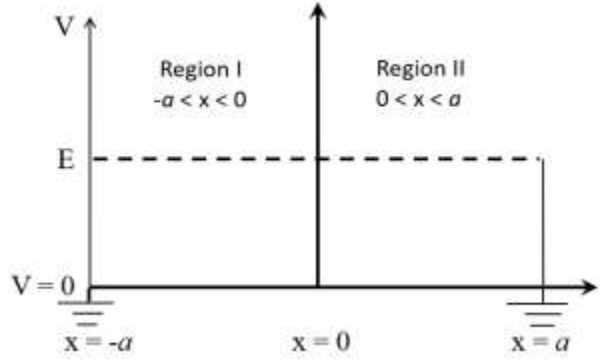

Fig. 10. Closed circuit with a single delta-function barrier.

We use the time-independent Schrödinger equation as Eq. 80 in regions I and II, which is modified in Eq. 81 at the origin where α is a constant.

$$-\frac{\hbar^2}{2m}\frac{d^2\psi}{dx^2}+(V-E)\psi=0 \tag{80}$$

$$-\frac{\hbar^2}{2m}\frac{d^2\psi}{dx^2}-\alpha\delta(x)\psi=E\psi \tag{81}$$

The coefficients for the wave function are not necessarily identical in the two half-spaces. Thus, we write the wavefunctions as Eqs. 82 and 84 in region I and the derivatives as Eqs. 83 and 85.

$$\psi_I(x)=A\cos(kx)+B\sin(kx) \tag{82}$$

$$\psi_{II}(x)=C\cos(kx)+D\sin(kx) \tag{83}$$

$$\psi_I{}'(x)=-kA\sin(kx)+kB\cos(kx) \tag{84}$$

$$\psi_{II}{}'(x)=-kC\sin(kx)+kD\cos(kx) \tag{85}$$

The wavefunctions at the two ends of the model are given by Eqns. 86 and 87, so continuity at these two ends which are connected requires that Eqn. 88 be satisfied.

$$\psi_I(-a)=A\cos(ka)-B\sin(ka) \tag{86}$$

$$\psi_{II}(a)=C\cos(ka)+D\sin(ka) \tag{87}$$

$$(A-C)\cos(ka)=(D+B)\sin(ka) \tag{88}$$

The derivative of the wavefunctions at both ends of the model are given by Eqns. 89 and 90, so continuity of the continuity of the wavefunctions at these two ends requires that Eqn. 91 be satisfied.

$$\psi_I{}'(-a)=kA\sin(ka)+kB\cos(ka) \tag{89}$$

$$\psi_{II}{}'(a)=-kC\sin(ka)+kD\cos(ka) \tag{90}$$

$$k(A+C)\sin(ka) = k(D-B)\cos(ka) \tag{91}$$

Thus, for Eqns. 88 and 91 to be satisfied when k not equal to zero, we must require that the following matrix equation is satisfied:

$$\begin{bmatrix} (A-C) & (D+B) \\ (D-B) & (A+C) \end{bmatrix} \begin{bmatrix} \cos(ka) \\ \sin(ka) \end{bmatrix} = \begin{bmatrix} 0 \\ 0 \end{bmatrix} \tag{92}$$

To have a unique solution for arbitrary values of the product ka it is necessary for the determinant of the matrix in Eq. 92 be zero as shown in Eq. 93. This requires that A, B, C, and D must satisfy Eq. 95.

$$\begin{vmatrix} (A-C) & (D+B) \\ (D-B) & (A+C) \end{vmatrix} = 0 \tag{93}$$

$$(A-C)(A+C) - (D+B)(D-B) = 0 \tag{94}$$

$$A^2 - C^2 - D^2 + B^2 = 0 \tag{95}$$

The wavefunction must also be continuous arbitrarily close to the origin so Eqs. 82 and 83 require that C is equal to A, and B is equal to D. We normalize by defining A to be unity so Eqs. 82 to 85 are simplified to give Eqs. 86 to 89, leaving only the two unknown coefficients B and D.

$$\psi_I(x) = \cos(kx) + B\sin(kx) \tag{86}$$

$$\psi_{II}(x) = \cos(kx) + D\sin(kx) \tag{87}$$

$$\psi_I'(x) = -k\sin(kx) + kB\cos(kx) \tag{88}$$

$$\psi_{II}'(x) = -k\sin(kx) + kD\cos(kx) \tag{89}$$

We also require continuity of the wavefunction and its derivative at the two ends of the model because these two points are connected which gives Eqs. 90 to 93 at these two points.

$$\psi_I(-a) = \cos(ka) - B\sin(ka) \tag{90}$$

$$\psi_{II}(a) = \cos(ka) + D\sin(ka) \tag{91}$$

$$\psi_I'(-a) = k\sin(ka) + kB\cos(ka) \tag{92}$$

$$\psi_{II}'(a) = -k\sin(ka) + kD\cos(ka) \tag{93}$$

Thus, for the wavefunction and its derivative to be consistent it is necessary to satisfy both Eqs. 94 and 95.

$$\cos(ka) - B\sin(ka) = \cos(ka) + D\sin(ka) \tag{94}$$

$$k\sin(ka) + kB\cos(ka) = -k\sin(ka) + kD\cos(ka) \tag{95}$$

Next, we rearrange Eq. 94 to obtain Eq. 96, and simplify Eq. 95 to obtain Eq. 97 so that now both Eq. 96 and Eq. 97 must be satisfied to obtain a solution.

$$(B+D)\sin(ka) = 0 \tag{96}$$

$$k\left[2\sin(ka) + (B-D)\cos(ka)\right] = 0 \tag{97}$$

If k were zero, then Eq. 96 and Eq. 97 would be satisfied for all values of the parameters a, B, and D. However, then the energy would be zero so this trivial case will not be considered. Thus, because we require that k is not zero, we divide Eq. 97 by k to obtain Eq. 98 where both Eq. 96 and Eq. 98 must be satisfied.

$$2\sin(ka) + (B-D)\cos(ka) = 0 \tag{98}$$

Equation 98 would be satisfied if B were equal to D and ka were nπ where n is an arbitrary integer. More generally we have Eq. 99 so that for each value of the product ka, the graph of D as a function of B is a single straight line with a slope of unity.

$$D = B + 2\tan(ka) \tag{99}$$

## VIII. POSSIBLE APPLICATIONS AT TERAHERTZ FREQUENCIES.

Here we consider the operation of a device under quasistatic conditions. Thus, if the size of the device were less than 10 nm the device could operate at frequencies up to about 1,000 THz, Then solutions of the time-dependent Schrödinger equation may be approximated by time-stepping a sequence of solutions of the time-independent Schrödinger equation. We consider the case for the rectangular barrier shown in Section II, which has thee four parameters a, b, E, and $V_0$. The expression for the determinant set equal to zero is given in Eq. 16.

In Section II, we specified any three of the four parameters in Eq. 16 and varied the fourth in order to bring the determinant to zero. Now we use Eq. 16 with Eq. 100 to obtain a quasi-static approximation for the time-dependent solution by the following five steps.

1. Specify $V_0$ and $V_1$. Typically, $V_1$ is much less than $V_0$ to avoid negative potentials.
2, Specify the value for ωt.
3, Use $V_0$, $V_1$, and ωt to calculate the potential V(ωt) using Eq. 100.

$$V_0(\omega t) \equiv V_0 + V_1 \sin(\omega t) \tag{100}$$

4. Vary the barrier length b to bring the determinant to zero for a solution.
5. Return to step 2 using other values of ωt to determine the solution over a specified range, such as for one period where the interval is from zero to 2π.

Note that b, the length of the barrier, is slightly greater than the length of the pre-barrier in each of the following four figures. Nonlinearity, caused by the variation of the barrier length in each cycle of the excitation, would cause the tunneling current to have harmonics at integer multiples of the excitation frequency. However, this effect would not be seen without using a model having greater precision.

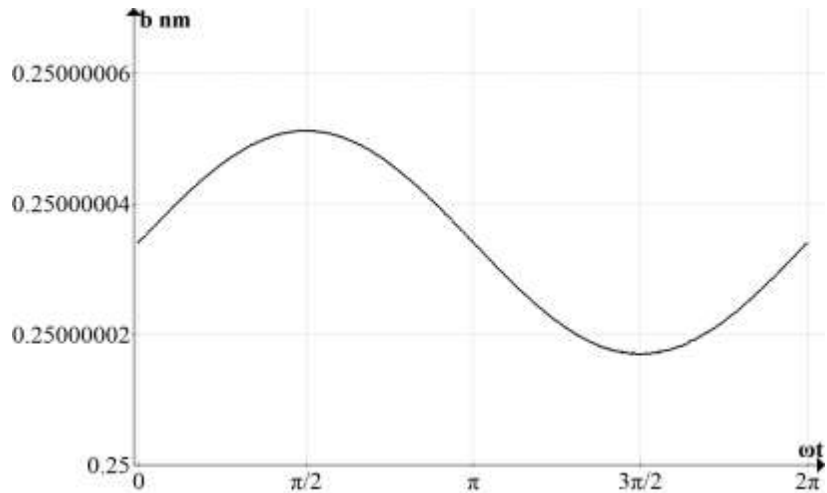

Fig 11. Barrier length b as a function of ωt for a pre-barrier length of 0.25 nm, with potential $V_0$ equal to 1 V and potential $V_1$ equal to 0.5 V in Eq. 100.

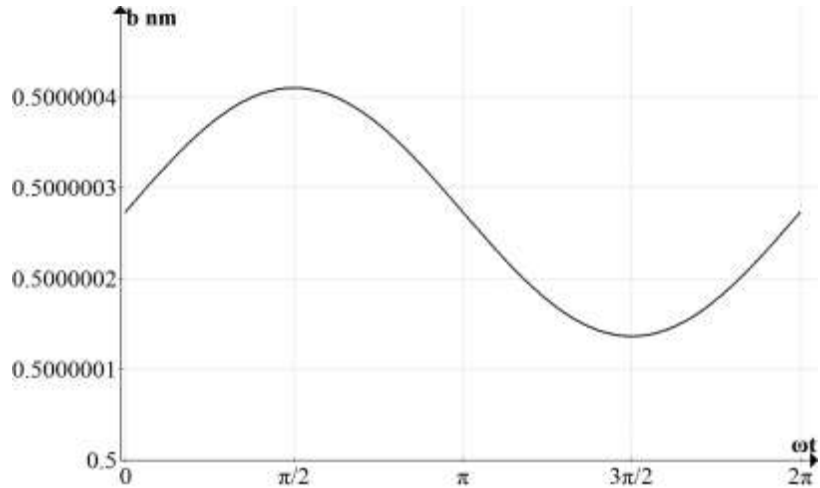
Fig 12. Barrier length b as a function of $\omega t$ for a pre-barrier length of 0.50 nm with potential $V_0$ equal to 1 V and potential $V_1$ equal to 0.5 V in Eq. 100.

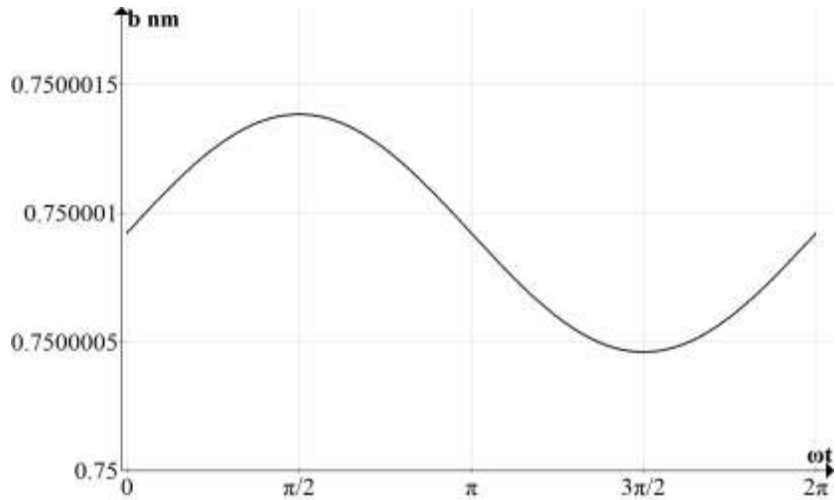
Fig 13. Barrier length b as a function of $\omega t$ for a pre-barrier length of 0.75 nm with potential $V_0$ equal to 1 V and potential $V_1$ equal to 0.5 V in Eq. 100.

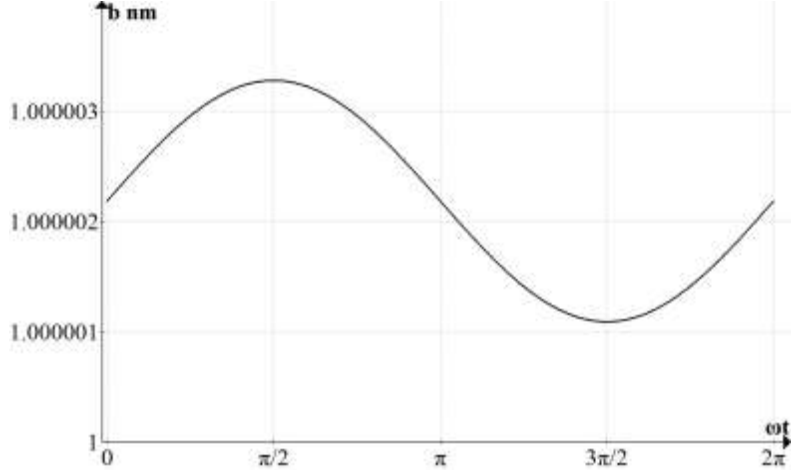

Fig 14. Barrier length b as a function of ωt for a pre-barrier length of 1.0 nm with potential $V_0$ equal to 1 V and potential $V_b$ equal to 0.5 V in Eq. 100.

Table I. Values of b, nm for the four figures.

| Figure | $V_1$ | Min | Mid | Max |
|---|---|---|---|---|
| 11 | 0.25 | 0.2500000170 | 0.2500000341 | 0.2500000512 |
| 12 | 0.50 | 0.5000001366 | 0.5000002734 | 0.5000004101 |
| 13 | 0.75 | 0.7500004614 | 0.7500009228 | 0.7500013842 |
| 14 | 1.00 | 1.0000010937 | 1.0000021874 | 1.0000032811 |

In the simulations that are shown in Figs. 11 to 14 the parameter ωt was varied through a full cycle as the pre-barrier length a was held constant. This combination was chosen because, in a possible related device, the pre-barrier would be a wire having a fixed length while the barrier would be a sub-nanometer gap which could be a gas or vacuum. Later simulations should be made using the triangular-barrier or other more realistic approximations before attempting to construct an actual device. Table I shows the values of the length of the tunneling junction in nanometers for each of the four figures. These values show a high degree of symmetry that would be expected with the small linear perturbation.

## X. DISCUSSION AND CONCLUSIONS

Conventional simulations for open-circuit models have an incident wave, a reflected wave, a transmitted wave, and two waves within the barrier. However, in each closed-circuit model, the four simultaneous equations are dependent so the determinant must be zero for a non-trivial solution. This results in groups of continuous solutions and/or singularities in the parameter space.

In each of the simulations that were made in this paper the determinant, which must be brought to zero, was initially on the order of 1,000. Iterations were continued using bisection until the values was less than $10^{-9}$.

The present set of examples is related to our earlier work at the Los Alamos National Laboratory as part of their CINT program (Center for Integrated Nanotechnologies). Then, we generated hundreds of microwave harmonics by focusing a mode-locked laser on the tunneling junction of a scanning tunneling microscope (STM) [5]. Each harmonic was at integer multiple of the laser pulse repetition rate (74.254 MHz). A bias-T in the tip circuit was used to measure harmonics superimposed on the DC tunneling current. The 200th harmonic at 14.85 GHz had a power of only 3.162 x10$^{-18}$ W. However, our analysis suggests that within the STM tunneling junction these harmonics extend to terahertz frequencies [6],[7]. The power measured at these harmonics fell off as the inverse square of their frequencies. This is consistent with the low-pass filter effect of the shunting capacitance and series resistance in the measurement circuit. We anticipate that this roll-off will be markedly reduced th the nanoscale devices described in the present paper because of the extremely small size for the entire circuit.

One possible implementation of this technology is what we call an "omega device" in which the wire loop corresponds to the upper section of a capital "Ω", the gap at the base corresponds to the tunneling junction. The two straight arms at the base of this symbol correspond to a dipole antenna receiving radiation from a mode-locked laser. A second, shorter, dipole at the base may be added to increase the transmission of the terahertz radiation. This is because an electrically-small circular loop has much lower radiation resistance than a dipole with comparable size. [8]. We attribute this loss to interference of the opposing fields at the two sides of the loop.

**ACKNOWLEDGMENT**

The author is grateful to our contractor Ezra Pedersen, who implemented our equations in this paper to perform nonlinear iterations with extremely large data sets to prepare some of the figures in this paper.